\documentclass[12pt,draftclsnofoot,journal,onecolumn]{IEEEtran}

\makeatletter
\long\def\@makecaption#1#2{\ifx\@captype\@IEEEtablestring%
	\footnotesize\begin{center}{\normalfont\footnotesize #1}\\
		{\normalfont\footnotesize\scshape #2}\end{center}%
	\@IEEEtablecaptionsepspace
	\else
	\@IEEEfigurecaptionsepspace
	\setbox\@tempboxa\hbox{\normalfont\footnotesize {#1.}~~ #2}%
	\ifdim \wd\@tempboxa >\hsize%
	\setbox\@tempboxa\hbox{\normalfont\footnotesize {#1.}~~ }%
	\parbox[t]{\hsize}{\normalfont\footnotesize \noindent\unhbox\@tempboxa#2}%
	\else
	\hbox to\hsize{\normalfont\footnotesize\hfil\box\@tempboxa\hfil}\fi\fi}
\makeatother

\usepackage{subfigure}
\usepackage{amsfonts}
\usepackage{amssymb}
\usepackage{cite}
\usepackage[cmex10]{amsmath}
\usepackage{float}
\usepackage{color}
\usepackage{amsthm}
\usepackage{lineno}
\usepackage{array}
\usepackage{makecell}
\usepackage{multirow}
\usepackage{tabu}
\usepackage{siunitx}
\usepackage[lined, boxed, linesnumbered, ruled]{algorithm2e}
\usepackage{algcompatible,amsmath}
\algnewcommand\algorithmicinput{\textbf{Input:}}
\algnewcommand\INPUT{\item[\algorithmicinput]}
\algnewcommand\algorithmicoutput{\textbf{Output:}}
\algnewcommand\OUTPUT{\item[\algorithmicoutput]}

\newtheorem{lemma}{Lemma}
\newtheorem{assumption}{Assumption}
\IEEEoverridecommandlockouts
\ifCLASSINFOpdf
\usepackage[pdftex]{graphicx}
\usepackage{epstopdf}
\DeclareGraphicsExtensions{.pdf,.jpeg,.png,.eps}
\else
\usepackage[dvips]{graphicx}
\DeclareGraphicsExtensions{.eps}
\fi

\usepackage{array}
\newcolumntype{P}[1]{>{\centering\arraybackslash}p{#1}}
\newcolumntype{M}[1]{>{\centering\arraybackslash}m{#1}}

\usepackage{verbatim}

\newtheorem{theorem}{Theorem}

\begin{document}
\title{Interference-Limited Ultra-Reliable and Low-Latency Communications: Graph Neural Networks or Stochastic Geometry?}

\author{Yuhong~Liu, Changyang~She, Yi~Zhong, Wibowo~Hardjawana, Fu-Chun Zheng, and Branka~Vucetic
\thanks{
	Yuhong Liu, Changyang She, Wibowo Hardjawana and Branka Vucetic are with School of Electrical and Information Engineering, The University of Sydney, Sydney, Australia. (Emails: yuhong.liu@sydney.edu.au, shechangyang@gmail.com, wibowo.hardjawana@sydney.edu.au, branka.vucetic@sydney.edu.au.)
	
	Yi Zhong is with School of Electronic Information and Communications, Huazhong University of Science and Technology, Wuhan, P. R. China. (Email: yzhong@hust.edu.cn.)
	
	Fu-Chun Zheng is with School of Electronic and Information Engineering, Harbin Institute of Technology (Shenzhen), Shenzhen, China (Email: fzheng@ieee.org.)
	
    Part of this work was presented at the 2019 IEEE Global Communications Conference (GLOBECOM) \cite{URLLC2019Wu}.

    }}
\maketitle

\begin{abstract}
In this paper, we aim to improve the Quality-of-Service (QoS) of Ultra-Reliability and Low-Latency Communications (URLLC) in interference-limited wireless networks. To obtain time diversity within the channel coherence time, we first put forward a random repetition scheme that randomizes the interference power. Then, we optimize the number of reserved slots and the number of repetitions for each packet to minimize the QoS violation probability, defined as the percentage of users that cannot achieve URLLC. We build a cascaded Random Edge Graph Neural Network (REGNN) to represent the repetition scheme and develop a model-free unsupervised learning method to train it. We analyze the QoS violation probability using stochastic geometry in a symmetric scenario and apply a model-based Exhaustive Search (ES) method to find the optimal solution. Simulation results show that in the symmetric scenario, the QoS violation probabilities achieved by the model-free learning method and the model-based ES method are nearly the same. In more general scenarios, the cascaded REGNN generalizes very well in wireless networks with different scales, network topologies, cell densities, and frequency reuse factors. It outperforms the model-based ES method in the presence of the model mismatch.
\end{abstract}

\section{Introduction}
The Fifth/Sixth Generation (5G/6G) of mobile networks are expected to support Ultra-Reliable and Low-Latency Communications (URLLC) for emerging applications, including industrial automation, intelligent transportation, multisensory mixed reality applications, and autonomous robotics \cite{Tutorial2021She, Latency2017Schulz, AVision2020Saad}. One of the major design objectives in 5G is to fulfill the two conflicting quality-of-service (QoS) requirements in terms of ultra-low latency and ultra-high reliability \cite{Elayoubi2019Radio}. The End-to-End (E2E) delay of URLLC is around $1$ ms, and the overall packet loss probability should be $10^{-5} \sim 10^{-7}$. In 6G, scalability is another challenge for URLLC, where the density of devices is high and the radio resources are limited \cite{Towards2019Ahmad}. To guarantee the QoS of URLLC, we should investigate the fundamental reliability-latency-scalability trade-off \cite{AVision2020Saad} and develop a flexible architecture that can be operated in large-scale networks \cite{Towards2019Ahmad}. 
 
Most of the existing studies on URLLC focused on single-cell scenarios, where inter-cell interference is not considered. They either optimized resource allocation to maximize the resource utilization efficiency subject to the QoS constraints of URLLC \cite{Resource2020Ghanem,Optimizing2019Sun,efffective2021Shehab,  ren2019joint,  Nasir2020MinMax, crosslayer2019she,yuan2021joint} or developed different kinds of retransmission/repetition schemes to improve the reliability in radio access networks  \cite{Resource2018Anand,3GPP2018NR}. Due to the limitation of radio resources, it is not possible to allocate orthogonal resource blocks to different users, especially in the scenarios with high device densities, such as smart factories and aerial/terrestrial vehicular networks. Motivated by this fact, resource sharing mechanisms were investigated in \cite{Librino2020Resource,liu2018d2dbased,Singh2018contentionbased,Mahmood2018Reliability}, where the impacts of intra-cell interference on the QoS of URLLC are studied. Nevertheless, in large-scale networks, inter-cell interference is the bottleneck of achieving URLLC. How to meet the QoS requirements of URLLC in the presence of inter-cell interference remains an open issue.

To analyze the inter-cell interference, the authors of \cite{Jiang2018Random,Lopez2018Aggregation,Qamar2019Stochastically, Harpreet2013Downlink,liu2020analyzing,zhong2017heterogeneous,zhong2020spatio} applied stochastic geometry to model the inter-cell interference in large-scale wireless networks and analyzed the relationship between transmission policies and some performance metrics such as successful transmission probability, coverage probability, and spectral efficiency. Since analytical tools in
stochastic geometry rely on uniform user distribution assumptions, machine learning methods were used to approximate interference distribution in \cite{Machine2020Shang,New2021Hmamouche}. 
The above methods are developed for performance analysis. They are inconvenient in policy optimization since one cannot derive closed-form expressions of latency and reliability when the topology and the resource allocation are complicated. As a result, novel methodologies for optimizing transmission policies for URLLC in interference-limited networks are much needed, especially when the network topology and status are dynamic. 

\subsection{Related Works}
The resource allocation of URLLC in a single-cell network has been well studied in the existing literature \cite{ Resource2020Ghanem, Optimizing2019Sun, ren2019joint,  Nasir2020MinMax, crosslayer2019she, yuan2021joint, efffective2021Shehab}. In \cite{Resource2020Ghanem}, the authors optimized the resource allocation for maximizing the weighted sum throughput subject to QoS constraints on the decoding error probability and the transmission delay. In \cite{Optimizing2019Sun}, the authors proposed a framework to maximize the energy efficiency with the latency and reliability constraints by optimizing the bandwidth and the power allocation. The authors in \cite{efffective2021Shehab} optimized the power allocation and the delay exponent to maximize the effective energy efficiency under the delay and reliability constraints. The works in \cite{ren2019joint} and \cite{Nasir2020MinMax} considered a joint power allocation and transmission blocklength optimization to minimize the decoding error probability with maximum delay and total power constraints. In \cite{crosslayer2019she}, a cross-layer design framework was proposed to minimize the overall packet loss probability subject to E2E delay requirements by optimizing the user association and bandwidth allocation. The results in \cite{yuan2021joint} indicate that using full-duplex relaying schemes makes it possible to achieve a better delay-reliability trade-off than the direct transmission. The authors in \cite{Resource2018Anand} analyzed the system capacity with given bandwidth, E2E delay, and delay violation probability. Further considering that the acknowledged feedback in retransmission schemes will cause extra delay in URLLC, the K-repetition scheme that does not require the acknowledged feedback between two consecutive repetitions was adopted in 5G New Radio (NR) \cite{3GPP2018NR}.

In interference-limited networks, the authors of \cite{Librino2020Resource} proposed an iterative algorithm that allocates the time and frequency resources to achieve URLLC, where a reliability threshold and a maximum tolerable delay constraint were considered. The authors in \cite{Elayoubi2019Radio} investigated resource allocation schemes for the initial transmission and retransmissions to meet the QoS requirements of URLLC. To avoid the request-and-grant procedure in uplink transmissions, a contention-based grant-free access mechanism was applied for URLLC in \cite{Singh2018contentionbased}. In \cite{liu2018d2dbased}, a novel two-phase transmission policy was optimized to maximize the total number of successful users in URLLC. The authors of \cite{Mahmood2018Reliability} designed a retransmission policy that transmits each packet multiple times through multiple links, as such the block error rate can be improved with spatial diversity. These studies provide valuable insights into designing URLLC systems with intra-cell interference. Nevertheless, the impact of inter-cell interference was not considered in these works. To analyze the inter-cell interference, stochastic geometry has been widely used in the existing literature. For example, the successful transmission probability \cite{Jiang2018Random,Lopez2018Aggregation} and the coverage probability \cite{Qamar2019Stochastically, Harpreet2013Downlink} have been obtained in interference-limited wireless networks. The analysis of wireless networks by combining stochastic geometry and queuing theory can be found in \cite{zhong2017heterogeneous,zhong2020spatio}. More recently, the authors of \cite{liu2020analyzing} analyzed the latency and reliability of different repetition schemes in interference-limited networks.

In the cases without closed-form analytical results, an artificial Neural Network (NN) can be used to approximate the optimal resource allocation policy that maps it to the QoS performance metric. As a universal approximator, fully-connected neural networks (FNN) has been applied in wireless communications to approximate complex policies, such as scheduling, power control, and bandwidth allocation. The dimensions of the input and the output of an FNN are determined before the training stage and cannot be adjusted according to the number of links in a wireless network. Thus, developing scalable neural networks in dynamic wireless networks becomes an urgent task. To tackle this issue, two kinds of Graph Neural Networks (GNN) were applied in \cite{Optimal2020MarkEisen} and \cite{Graph2021Shen}, respectively. The goal was to develop flexible GNNs for maximizing the sum rate of a wireless network. Since these existing solutions are not developed for URLLC, how to meet the stringent latency and reliability requirements in interference-limited networks deserves further study.

\subsection{Our Contributions}
In this paper, we aim to minimize the QoS violation probability of URLLC in an interference-limited wireless network, which is defined as the percentage of users that cannot achieve URLLC. If the packet loss probability of a user is larger than a required threshold, the QoS of the user is violated. The major contributions of this paper are summarized as follows.
\begin{itemize}
\item We put forward a random repetition policy in the time domain. Like the K-repetition scheme in 5G NR \cite{3GPP2018NR}, our repetition policy reserves multiple time slots for each packet and does not require acknowledgment feedback between two consecutive transmissions. The difference is that the transmitter randomly chooses some of the time slots and sends a copy of the packet in each selected time slot. In the remainder of the time slots, the transmitter remains silent. As a result, the interference power from the transmitter to other receivers is random in different time slots, depending on whether the transmitter is active or not. This policy can reduce the QoS violation probability by exploiting the variation of interference in interference-limited networks.

\item We establish an optimization framework by incorporating communication, queuing, and network models to find the optimal repetition policy. The E2E delay consists of the queuing delay in the buffer of the transmitter and the repetition delay in the wireless network. The queuing delay violation probability and the decoding error probability in the short coding blocklength regime are considered in the overall packet loss probability.

\item
We build a cascaded random-edge graph neural network (REGNN) to find the optimal repetition scheme in an arbitrary wireless network. Considering the number of repetitions should not exceed the number of time slots, the first REGNN outputs the number of time slots for all the links; these are the input of the second REGNN that outputs the number of repetitions. Due to the complicated communication, queueing, and network models, the expression of the QoS violation probability is not available in general. To overcome this difficulty, we develop a model-free unsupervised learning algorithm, and derive the gradients of the loss function (without an explicit expression) with respect to the parameters in the two REGNNs, respectively.

\item
To obtain some analytical results, we simplify the network model by considering a symmetric scenario and derive the approximations of the QoS violation probability using stochastic geometry. Specifically, we assume the packet arrival rates and the large-scale channel gains of all the links are the same in a Poisson bipolar network. With the help of these assumptions, we can simplify the optimization problem and find the optimal repetition policy by Exhaustive Search (ES). The policy obtained from model-based ES in this symmetric scenario can serve as a benchmark to validate the effectiveness of the model-free learning method.

\item
The simulation results show that our proposed cascaded REGNN can achieve a near-optimal solution in symmetric scenarios and can generalize well in more general scenarios. A well-trained cascaded REGNN can be applied to interference-limited networks with different numbers of links, user densities, and network topologies. Although it is possible to obtain analytical results with stochastic geometry, the model-based ES method degrades in more general scenarios due to model mismatch.
\end{itemize}

The remaining part of the paper is organized as follows. Section II describes the system model and the problem formulation. Section III introduces the structure of the cascaded REGNN and the unsupervised learning algorithm. Section IV derives the bounds of the QoS violation probability in a symmetric scenario. Section V shows the numerical and simulation results. Finally, Section VI concludes this paper.

\section{System Model and Problem Formulation}\label{system_model}

\subsection{Network and Channel Model}
\begin{figure}[htbp]
\label{fig:sys model}
\centering 
\subfigure[]{                    
\begin{minipage}{5cm}
\centering                                                  
\includegraphics[scale=0.5]{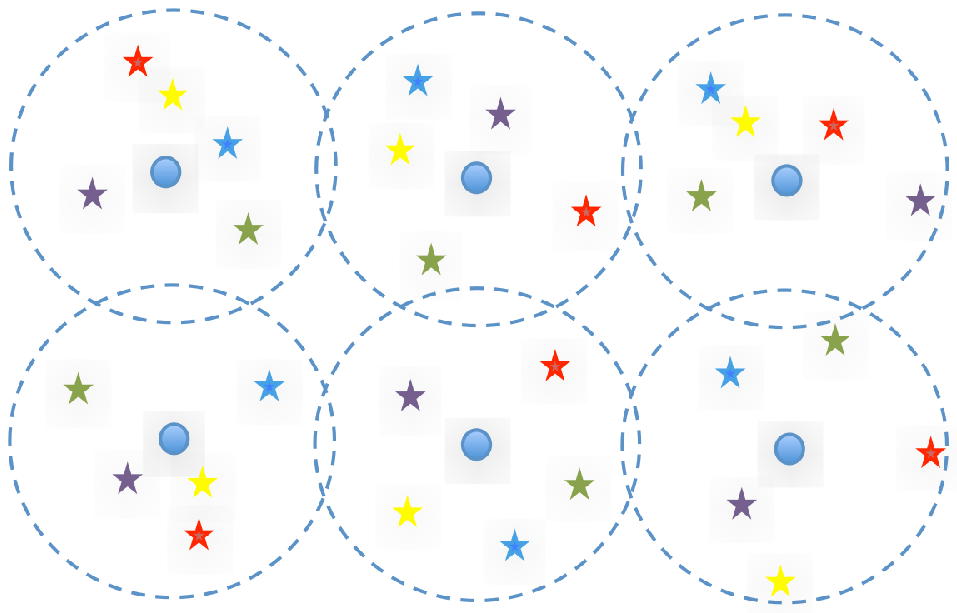}
\label{fig:system_a}
\end{minipage}}
\subfigure[]{                   
\begin{minipage}{5cm}
\centering                                                  
\includegraphics[scale=0.425]{ 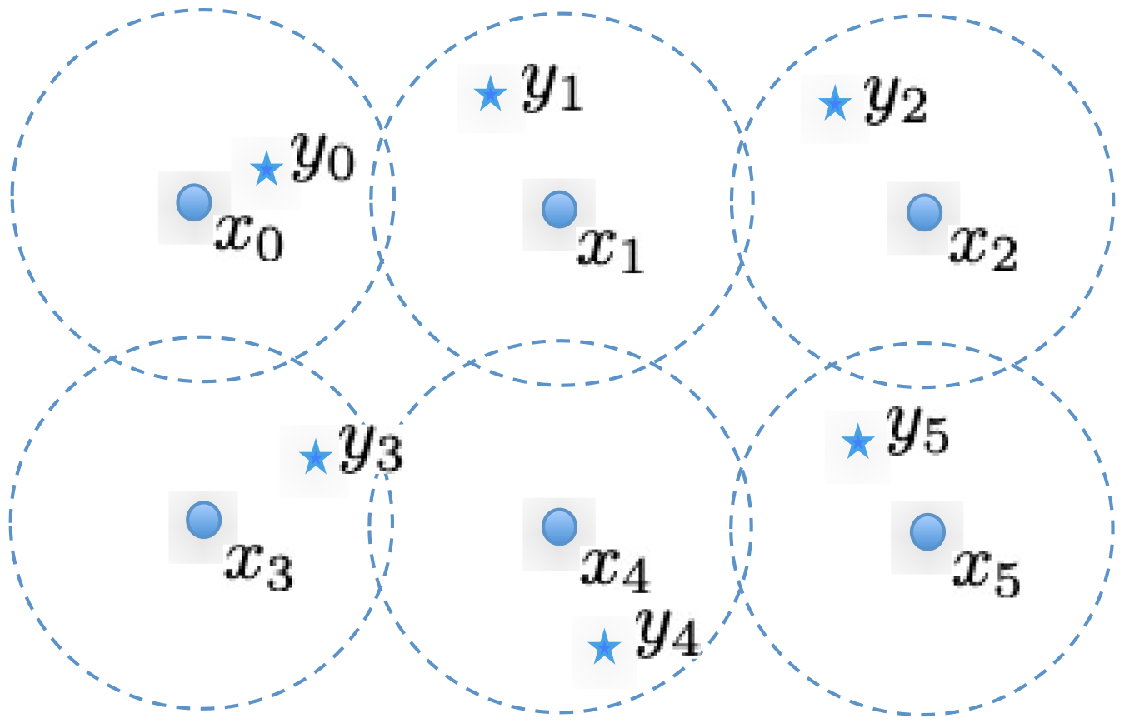}
\label{fig:system_b}
\end{minipage}}
\subfigure[]{                   
\begin{minipage}{5cm}
\centering                                                 
\includegraphics[scale=0.425]{ 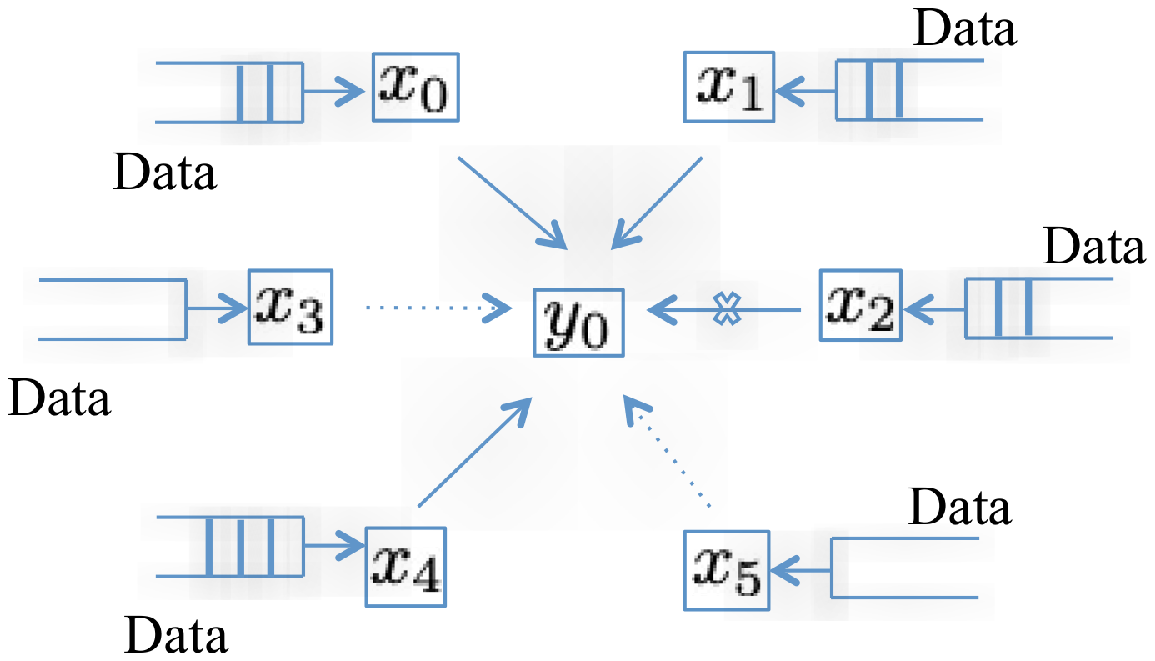}
\label{fig:system_c}
\end{minipage}}
\caption{A snapshot of the network model. Fig. \ref{fig:system_a} shows a cellular network with OFDMA. The circles represent the BSs, and the pentagrams represent the users. Users with different colors use different subchannels. Fig. \ref{fig:system_b} shows the users using the same subchannel. Fig. \ref{fig:system_c} illustrates the interference of one link in the considered time slot. The solid arrows represent the transmitters that are active and transmitting packets in the current time slot. If there is a cross on the solid arrow, it means the transmitter has a packet in its buffer but it stays silent with the random repetition. The dot arrows mean there is no packet in the buffer of the transmitter, and hence there is no interference from this transmitter.}

\end{figure}

Consider a wireless network shown in Fig. \ref{fig:system_a}, where the network topology could be modeled as a hexagonal grid, a Poisson bipolar network, or other models. Each base station equipped with $N_{\rm T}$ antennas serves multiple single-antenna users with orthogonal frequency division multiple access (OFDMA) to avoid intra-cell interference. The frequency-reuse factor could be $1$, $1/3$ or $1/7$ and is selected based on the time and frequency resources and the network topology. We consider a massive URLLC scenario, where the inter-cell interference is not negligible even when the frequency reuse factor is less than one. We assume that the interference on different subchannels is independent and identically distributed, and analyze the interference on one subchannel of the OFDMA system as shown in Fig. \ref{fig:system_b}. The number of wireless links (BS-user pairs) could be a random variable or a fixed value according to the network model. Denote the number of wireless links (BS-user pairs) in the network using the same subchannel and interfering with each other as $K$. The $k$-th BS-user pair is referred to as the $k$-th link in the sequel.  

Time is discretized into slots. The duration of each slot is denoted by $T_{\rm{s}}$. The transmitter either has a packet to transmit or stays silent in each time slot. Given a stringent latency requirement, we can assume that each transmitter has a buffer with sufficient capacity to store the data packets. This is because the packet losses are caused by delay bound violation, not buffer overflow. The arrival processes of packets at different transmitters are independent and follow Poisson processes with arrival rates $\boldsymbol{\lambda} \triangleq [\lambda_1,\lambda_2,...,\lambda_K]^T$ (packets/slot). To decrease the interference from other links and improve the reliability, we propose a random repetition policy that allows the transmitter to send multiple copies of each packet without waiting for the acknowledgment feedback from the receiver. Specifically, the $k$-th transmitter transmits one packet by using $N_k$ time slots. To obtain time diversity (i.e., interference varies in different slots \cite{URLLC2019Wu}), $M_k \in [1,N_k]$ slots are randomly selected from $N_k$ time slots to send the same packet for $M_k$ times. In the other $N_k-M_k$ slots, the transmitter remains silent. In this way, the network benefits from the diversity of interference in different slots.

Let the large-scale channel gain and the small-scale channel coefficient of the $k$-th link be $\mu_{kk}$ and $\textbf{h}_{kk} \triangleq [h_{kk}^1, h_{kk}^2,...,h_{kk}^{N_{\rm{T}}}]^T$, respectively. We consider the block fading channel model, where the small-scale channel coefficients remain constant within each block with duration $T_{\rm{c}}$ (i.e., the channel coherence time) and vary independently among different blocks. For low-latency communications, the required delay bound is shorter than the channel coherence time \cite{Tutorial2021She}. Thus, the channel is quasi-static in each slot. We assume that each link's channel state information (CSI) is only available at the receiver. Based on the received signal in \cite[Chapter~5.4.3]{FundamentalsOfWC}, we can obtain the signal-to-interference-and-noise-ratio (SINR) of the $k$-th link at the $t$-th time slot, i.e.,
\begin{equation}\label{SINR}
	\gamma_k(t)=\frac{p_k \mu_{kk}	 \textbf{h}_{kk}^*\textbf{h}_{kk}/{N_{\rm T}}}
 {\sum_{i=1,i\neq k}^{K}(p_i \mu_{ik} {\textbf{h}_{ik}^*\textbf{h}_{ik}}/{N_{\rm T}})
	 \textbf{1}_i(t) + \sigma^2},
\end{equation}
where $\mu_{ik}$ and $\textbf{h}_{ik}$ are the large-scale channel gain and small-scale channel coefficients from the transmitter of the $i$-th link to the receiver of the $k$-th link, respectively. $p_k$ is the transmit power of the $k$-th transmitter. $\textbf{1}_i(t)$ is the indicator that shows the transmission state of $i$-th link at time slot $t$. If the $i$-th transmitter is active, i.e., sending a packet in the $t$-th slot, then $\textbf{1}_i(t) = 1$. Otherwise, $\textbf{1}_i(t) = 0$. Since each packet is transmitted for $M_i$ times, given the packet arrival rate $\lambda_i$, the probability that the $i$-th transmitter is active is ${\mathbb{E}}\{\textbf{1}_i(t)\} = \lambda_i M_i$. $\sigma^2$ is noise power at the receiver. 

\subsection{QoS Formulation}
The quality-of-service (QoS) requirement of each user is characterized by the E2E delay and the overall packet loss probability. The overall packet loss probability includes a decoding error probability at the receiver and a queuing delay violation probability at the transmitter. We analyze the above mentioned two probabilities under the E2E delay requirement below.
\subsubsection{Decoding error probability} Denote the decoding error probability of the packet transmitted in the $t$-th slot by $\epsilon_k^{\rm{d}}(t)$. The number of bits in each packet is denoted by $a$. According to the analysis in \cite{Scarlett2017Dispersion}, if the ratio of the fourth moment of the interference plus noise to the second moment of it is higher than $3$, the achievable rate can be approximated by \cite{Yury2010Channel}
\begin{equation} \label{R}
R_{k}\approx \frac{T_{\rm{s}} W}{a \ln2} \bigg[\ln (1+  \gamma_k(t)) -\sqrt{ \frac{V_{k}} {T_{\rm{s}} W} } f_{\rm{Q}}^{-1}(\epsilon_k^{\rm{d}}(t)) \bigg], ~~(\text{packets/slot})
\end{equation}
where $W$ is the bandwidth of the subchannel, $f_{\rm{Q}}^{-1}$ is the inverse of the Q-function, and $V_{k}=1-[1+ \gamma_k ]^{-2}$. Given some assumptions on the codewords, we can derive more accurate approximations of $R_k$. The numerical results in \cite{Scarlett2017Dispersion} indicate that as the number of interference links increases, the gaps between different approximations decrease. Therefore, using different approximations will not lead to fundamental differences in massive URLLC scenarios.

To transmit one packet in one slot ($R_k=1$), we can derive the decoding error probability in the $t$-th slot as follows,
\begin{equation}\label{decodingErrorU}
\epsilon_k^{\rm{d}}(t)
= f_{\rm{Q}} \left\{ \sqrt{\frac{T_{\rm{s}} W}{V_k}} \left[ \ln(1+ \gamma_k(t))-\frac{a \ln 2}{T_{\rm{s}} W} \right] \right\}.
\end{equation}
Let's denote the packet loss probability with $M_k$ repetitions by $\epsilon_k^{\rm{c}}$. The relationship between $\epsilon_k^{\rm{d}}(t)$ and $\epsilon_k^{\rm{c}}$ is given by
\begin{equation}\label{dep}
\epsilon_k^{\rm{c}}= \prod_{t=1}^{M_k}\epsilon_k^{\rm{d}}(t).
\end{equation}

\subsubsection{Queuing delay violation probability} Let the E2E delay bound be $D_{\max}$ (slots), including the queuing delay $D_k^{\rm q}$ and the transmission delay $D_k^{\rm t}$. Given $N_k$, the transmission delay is $D_k^{\rm t} = N_k$. To meet the E2E delay requirement, the following constraint should be satisfied,
\begin{equation} \label{eq_u_dr}
D_k^{\rm{q}}+D_k^{\rm{t}} \leq D_{\max}.
\end{equation}
The maximum queuing delay bound is given by $D_{k,\max}^{\rm q} = D_{\max} - N_k$. Considering that the queuing delay is a random variable, we denote the queuing delay violation probability by $\epsilon_k^{\rm{q}} \triangleq \mathbb{P}(D_k^{\rm q} > D_{k,\max}^{\rm q} )$.

For the stochastic packet arrival process at each transmitter, if the peak packet arrival rate is higher than the service rate, then there is a queue in the buffer of the transmitter. To characterize the queuing delay violation probability, $\epsilon_k^{\rm q}$, we use effective bandwidth, which is defined as the minimal constant service rate that is required to meet the target queuing delay requirement \cite{EB}. For the Poisson process, the effective bandwidth for $k$-th transmitter is given by \cite{Cross2018she}
\begin{equation}\label{EffectiveBW}
	E_k^{\rm{B}}(\theta_k)=\frac{\lambda_k}{\theta_k} (e^{\theta_k}-1),~~\text{(packets/slot)}
\end{equation}
where $\theta_k \triangleq \ln [ \frac{T_{\rm{s}}\ln(1/\epsilon_k^{\rm{q}})}{\lambda_k D_{k,\max}^{\rm{q}}}+1 ]$ depends on the delay bound, delay bound violation probability, and packet arrival rate. With the repetition policy, the service time of each packet is $N_k$ time slots. Thus, the service rate is constant, $1/N_k$. According to the definition of effective bandwidth, we can obtain the queuing delay violation from the following expression,
\begin{align}\label{eq:dvp_con}
    E_k^{\rm{B}}(\theta_k) = 1/N_k.
\end{align}
Since effective bandwidth decreases with $N_k$, given the queuing delay bound $D_{k,\max}^{\rm{q}}$ and the delay bound violation probability $\epsilon_k^{\rm{q}}$, the value of $N_k$ can be obtained via binary search.

\subsubsection{Overall packet loss probability and the QoS violation probability} Based on the previous analysis, the overall packet loss probability of the $k$-th link is given by
\begin{align} \label{eq: total_error}
1-(1-\epsilon_k^{\rm{q}})(1-\epsilon_k^{\rm{c}}) \approx \epsilon_k^{\rm{q}} + \epsilon_k^{\rm{c}}.
\end{align}
Given the maximum tolerable packet loss probability, $\epsilon_k^{\max}$, the probability that the delay and reliability requirement of the $k$-th link cannot be satisfied is given by
\begin{equation} \label{eq:P_k}
P_k = \mathbb{P} (\epsilon_k^{\rm{q}}+\epsilon_k^{\rm{c}} > \epsilon_k^{\max}),~ k \in K.
\end{equation}

We define the QoS violation probability of the network as the percentage of links that cannot achieve URLLC, i.e.,
\begin{equation}\label{eq:P_a}
{{P}_{{\rm vio}}} = \frac{1}{K} \sum_k^K \left[ 1 \times P_k + 0 \times (1-P_k) \right ] = \frac{1}{K} \sum_k^K P_k, k \in K.
\end{equation}

\subsection{Problem Formulation}\label{sec:problem formulation}
Based on the previous analysis, we can formulate an optimization problem to minimize the QoS violation probability by optimizing the number of slots reserved for each packet, $\textbf{N}(\textbf{H},\boldsymbol{\lambda})=[N_1,N_2,...,N_K]^T$, and the number of repetitions, $\textbf{M}(\textbf{H},\textbf{N})=[M_1,M_2,...,M_{K}]^T$, where $\textbf{H}$ is the matrix of the large-scale channel gains, defined as
\begin{equation}
\textbf{H} \triangleq \begin{bmatrix}
\mu_{1,1}&\mu_{1,2}&\dots&\mu_{1,K} \\
&&\dots&\\	
\mu_{K,1}&\mu_{K,2}&\dots&\mu_{K,K}
\end{bmatrix}.
\end{equation}
Since the small-scale channel coefficients vary every few milliseconds, collecting instantaneous channel gains will lead to an extremely high communication overhead. Therefore, we assume that only the large-scale channel gains are available for optimizing the repetition scheme. The problem can be formulated as follows,
\begin{align}\label{op}
\min_{\textbf{N}(\textbf{H},\boldsymbol{\lambda}), ~\textbf{M}(\textbf{H},\textbf{N})} 
  &~~~
  P_{{\rm vio}} = \left[\frac{1}{K}\sum_{k=1}^{K} \mathbb{P}(\epsilon_k^{\rm{q}}+\epsilon_k^{\rm{c}} > \epsilon_k^{\max}) \right]\\
 \text{s.t} ~~
  & 1 \leq N_k \leq N_k^{{\rm q},\max}, k=1,2,...,K, \label{eq:stable} \tag{\theequation a}\\
  & 1\leq M_k \leq N_k, k=1,2,...,K, \label{eq:activeT} \tag{\theequation b}\\
   & \eqref{dep}, ~\eqref{eq:dvp_con},\nonumber 
\end{align}
where $\textbf{N}$ and $\textbf{M}$ are integers. The maximum number of slots reserved to each packet in \eqref{eq:stable} is defined as $N_k^{{\rm q},\max}$, which is the maximum number of slots that can meet the queuing delay violation probability, $\epsilon_k^{\rm q}\leq\epsilon_k^{\max}$. Since effective bandwidth in \eqref{eq:dvp_con} decreases with $N_k$, we can find $N_k^{{\rm q},\max}$ via binary search. Constraint \eqref{eq:activeT} guarantees that the number of repetitions does not exceed the number of slots reserved to each packet. 

Given the realizations of large-scale channel gains and average packet arrival rates, the above problem is a non-linear integer programming. Besides, the QoS violation probability does not have a closed-form expression. Thus, finding the solution to the above problem is very challenging. Further considering that the large-scale channel gains, average packet arrival rates, and the number of links are dynamic, we need to update the optimal transmission policy according to these dynamic parameters in real-time. In the next two sections, we illustrate how to solve this problem by using a model-free unsupervised learning method and a model-based ES method, respectively.

\section{Model-Free Unsupervised Learning: A GNN Approach}
Although it is possible to use optimization techniques to find the optimal repetition policy for specific realizations of traffic state $\boldsymbol{\lambda}$ and network state $\textbf{H}$, the system needs to execute the optimization technique whenever the realizations of $\boldsymbol{\lambda}$ and $\textbf{H}$ change. The computation complexity of the optimization algorithm can be very high when the number of links in the network is large. To develop a solution with low complexity, we use GNN as a function approximator. The GNN maps the traffic state $\boldsymbol{\lambda}$ and the network state $\textbf{H}$ to the optimal repetition policy. We train the parameters of the GNN by unsupervised deep learning. After the training stage, we can use the forward propagation algorithm to obtain the optimal repetition policy for any realizations of traffic state and network state. 

Before we set up our GNN, it is worth noting that the gradient of the loss function with respect to the optimization variables is small when $\mathbb{P}(\epsilon^{\rm q}_k+\epsilon^{\rm c}_k > \epsilon^{\max}_k)$ is close to zero. To improve the learning efficiency, we turn to minimizing the logarithmic value of QoS violation probability. The optimization problem in \eqref{op} can be re-expressed as follows, 
\begin{align}\label{op_in_training}
\min_{\textbf{N}(\textbf{H},\boldsymbol{\lambda}), ~\textbf{M}(\textbf{H},\textbf{N})} 
  &~~~
  \log \left[\frac{1}{K}\sum_{k=1}^{K} \mathbb{P}(\epsilon_k^{\rm{q}}+\epsilon_k^{\rm{c}} > \epsilon_k^{\max}) \right]\\
 \text{s.t} ~~
  & 1 \leq N_k \leq N_k^{{\rm q},\max}, k=1,2,...,K, \label{eq:stable_train} \tag{\theequation a}\\
  & 1\leq M_k \leq N_k, k=1,2,...,K, \label{eq:activeT_train} \tag{\theequation b}\\
   & \eqref{dep}, ~\eqref{eq:dvp_con}.\nonumber 
\end{align}

\subsection{Cascaded Random Edge Graph Neural Network}
Although fully connected neural networks (FNN) have been widely applied in functional approximation, it is not suitable for dynamic wireless networks since the number of parameters to be trained (e.g., the number of neurons) depends on the number of devices that are changing over time. Compared with the fully connected neural network, another kind of structure, called Random Edge Graph Neural Network (REGNN), is more suitable for dynamic wireless networks, as shown in \cite{Optimal2020MarkEisen}. The number of nodes in the graph equals the number of wireless links in the network. Each link is represented by a node. The topology of the interference network is represented by the random matrix $\textbf{H}$. In other words, the edge from node $i$ to node $j$ is the large-scale channel gain from the transmitter of the $i$-th link to the receiver of the $j$-th link. With a graph convolutional architecture, the number of parameters in a REGNN does not change with the input and output dimensions. Thus, a well-trained REGNN can be applied to interference networks with a different number of links.

\begin{figure}[h]
	\centering
	\includegraphics[width=0.5\textwidth]{ 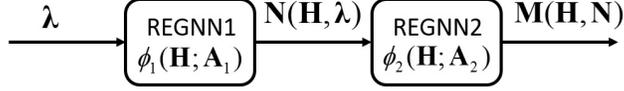}
	\caption{Cascaded Random Edge Graph Neural Network}\label{Fig:C_GNN}
\end{figure}

To find the optimal repetition policy in dynamic wireless networks, we represent the repetition policy, $\textbf{N}(\textbf{H},\boldsymbol{\lambda})$ and $\textbf{M}(\textbf{H},\textbf{N})$, by two REGNNs. Considering the input of $\textbf{M}(\textbf{H},\textbf{N})$ depends on the output of $\textbf{N}(\textbf{H},\boldsymbol{\lambda})$, we develop a cascaded REGNN structure as shown in Fig. \ref{Fig:C_GNN}. Denote the operations and the trainable parameters of the two REGNNs by $\boldsymbol{\phi}_{j}$ and $\textbf{A}_{j}$, $j\in \{1,2\}$, we have
\begin{align}
&\textbf{N}(\textbf{H},\boldsymbol{\lambda}) = \boldsymbol{\phi}_1 (\textbf{H};\textbf{A}_1)\boldsymbol{\lambda}, \label{c-regnn-1}
\\
&\textbf{M}(\textbf{H},\textbf{N}) = \boldsymbol{\phi}_{2}(\textbf{H};\textbf{A}_2)\textbf{N}(\textbf{H},\boldsymbol{\lambda}) = \boldsymbol{\phi}_{2}(\textbf{H};\textbf{A}_2)\boldsymbol{\phi}_1 (\textbf{H};\textbf{A}_1)\boldsymbol{\lambda} \label{c-regnn-2}.%
\end{align}
The operations include graph filters, activation functions, and aggregations of multiple graph filters in the hidden layers. The trainable parameters of the graph contain multiple sets of filter weights. To be more specific, if we have $F^{[l]}_j$ graph filters in the $l$-th hidden layer, the output of $l$-th layer is given by \cite{Graph2021Shen},
\begin{equation}
\begin{aligned}
	\textbf{y}^{[l]}_j = \psi_j^{[l]}\left(\sum_{f=1}^{F^{[l]}_j}\sum_{i=0}^{I^{[l]}_j}A_{i,j}^{[l,f]} \textbf{H}^{i} \textbf{y}^{[l-1]}_j\right),
\end{aligned}
\end{equation}
where $\textbf{y}^{[l]}_j$ is the output of the $l$-th layer (i.e., input of the $(l+1)$-th layer), $\psi_j^{[l]}$ is the point-wise nonlinear activation function, $I^{[l]}_j$ is the length of the filter, $A_{i,j}^{[l,f]}, i = 0,1,...,I^{[l]}_j$, are the coefficients of the $f$-th filter, and $\textbf{H}^{i}$ is the $i$-th power of the square matrix $\textbf{H}$. With total number of $L_j$ layers, the trainable parameters of $j$-th REGNN are $\textbf{A}_j = \{A_{i,j}^{[l,f]}\},i = 0,1,..., I^{[l]}_j, f=1,2,...,F_j^{[l]}$, and $l = 1,2,..., L_j$.

\subsection{Training of Cascaded REGNN}\label{Training_steps}
The cascaded REGNN can be trained iteratively by using the stochastic gradient descent algorithm. To apply the stochastic gradient descent algorithm, we need the gradient of the loss function with respect to (w.r.t) the trainable parameters. The chain rule of composition functions is widely used to compute the gradient and it needs the gradients of the loss function w.r.t. $\boldsymbol{\phi}_1(\textbf{H};\textbf{A}_1)$ and $\boldsymbol{\phi}_2(\textbf{H};\textbf{A}_2)$, and the gradients of $\boldsymbol{\phi}_1(\textbf{H};\textbf{A}_1)$ and $\boldsymbol{\phi}_2(\textbf{H};\textbf{A}_2)$ w.r.t. $\textbf{A}_1$ and $\textbf{A}_2$. Since the loss function in our problem does not have closed-form expression, such an approach is not feasible. To overcome this difficulty, we apply the model-free learning method in \cite{Learning2019MarkEisen}. This learning method estimates the gradient without the explicit expression of the loss function. This is done by replacing the deterministic policies $\textbf{N}(\textbf{H},\boldsymbol{\lambda})$ and $\textbf{M}(\textbf{H},\textbf{N})$ with two stochastic policies $\hat{\textbf{N}}$ and $\hat{\textbf{M}}$. The policy $\hat{\textbf{N}}$ depends on the trainable parameters of the first REGNN $\textbf{A}_1$ and the input realization of $\textbf{H}$ and $\boldsymbol{\lambda}$, so we denote the probability density function (pdf) of $\hat{\textbf{N}}$ as $\Psi_{\hat{\textbf{N}}}(\textbf{z}_1;\textbf{A}_1|\textbf{H},\boldsymbol{\lambda})$. $\textbf{z}_1$ is a vector of random variables generated from $\Psi_{\hat{\textbf{N}}}(\textbf{z}_1;\textbf{A}_1|\textbf{H},\boldsymbol{\lambda})$. As the policy $\hat{\textbf{M}}$ depends on the trainable parameters of the second REGNN $\textbf{A}_2$ and the input realization of $\textbf{H}$ and $\textbf{z}_1$, we denote the conditional pdf of the policy $\hat{\textbf{M}}$ as $\Psi_{\hat{\textbf{M}}|\hat{\textbf{N}}}(\textbf{z}_2;\textbf{A}_2|\textbf{H},\textbf{z}_1)$. $\textbf{z}_2$ is a vector of random variables generated from $\Psi_{\hat{\textbf{M}}|\hat{\textbf{N}}}(\textbf{z}_2;\textbf{A}_2|\textbf{H},\textbf{z}_1)$. If the density functions of the stochastic policies are represented by two impulse functions, i.e., $\Psi_{\hat{\textbf{N}}}(\textbf{z}_1;\textbf{A}_1|\textbf{H},\boldsymbol{\lambda})=\delta(\textbf{z}_1-
\textbf{N}(\textbf{H},\boldsymbol{\lambda}))$, $\Psi_{\hat{\textbf{M}}|\hat{\textbf{N}}}(\textbf{z}_2;\textbf{A}_2|\textbf{H},\textbf{z}_1)=\delta(\textbf{z}_2-\textbf{M}(\textbf{H},\textbf{N}))$, then the stochastic policies are exactly the same as the deterministic policies. In this case, the gradient can be calculate based on the policy gradient method from \cite{Learning2019MarkEisen},
\begin{align}
    &\nabla_{\textbf{A}_{1}} \mathbb{E}_{\boldsymbol{\lambda},\textbf{H},\textbf{N},\textbf{M}} f_{\rm loss} \left ( \boldsymbol{\lambda},\textbf{H},\textbf{N}, \textbf{M} \right ) \nonumber\\
    & =    \mathbb{E}_{\boldsymbol{\lambda},\textbf{H},\textbf{z}_1,\textbf{z}_2} \left \{ f_{\rm loss}(\boldsymbol{\lambda},\textbf{H},\textbf{z}_1,\textbf{z}_2) \nabla_{\textbf{A}_{1}} [\log  (\Psi_{\hat{\textbf{N}}}(\textbf{z}_1;\textbf{A}_1|\textbf{H},\boldsymbol{\lambda})
     ] \right \}, \label{eq:gradientEstimation1} \\ 
    &\nabla_{\textbf{A}_{2}} \mathbb{E}_{\boldsymbol{\lambda},\textbf{H},\textbf{N},\textbf{M}} f_{\rm loss} \left ( \boldsymbol{\lambda},\textbf{H},\textbf{N}, \textbf{M} \right ) \nonumber\\
    &= \mathbb{E}_{\boldsymbol{\lambda},\textbf{H},\textbf{z}_1,\textbf{z}_2} \left \{ f_{\rm loss}(\boldsymbol{\lambda},\textbf{H},\textbf{z}_1,\textbf{z}_2) \nabla_{\textbf{A}_{2}} [\log ( \Psi_{\hat{\textbf{M}}|\hat{\textbf{N}}}(\textbf{z}_2;\textbf{A}_2|\textbf{H},\textbf{z}_1))] \right \} 
    \label{eq:gradientEstimation2}
\end{align}
where $f_{\rm loss}(.)$ can be arbitrary loss function that does not have closed-form expression. The derivations of \eqref{eq:gradientEstimation1} and \eqref{eq:gradientEstimation2} are provided in Appendix A. 

To compute the right-hand sides of \eqref{eq:gradientEstimation1} and \eqref{eq:gradientEstimation2}, we can use a differentiable distribution function to approach the impulse function, such as truncated Gaussian distribution. The $j$-th REGNNs outputs the parameters of the distribution functions for the $k$-th user, denoted by $(\xi_j(k),\beta_j(k))$. Then, the values of $\textbf{z}_1$ and $\textbf{z}_2$ are generated randomly according to these distribution functions. By training the REGNNs, we update $(\xi_j(k),\beta_j(k))$ such that the distribution functions can approach impulse functions.

To train the cascaded REGNN, the system needs to execute the following five steps iteratively. We use a superscript $x^{\{\tau \}}$ to represent the value of a parameter in the $\tau$-th iteration. The algorithm is summarized in Algorithm \ref{alg:A1}.

\subsubsection{Initialization} We first select proper values of hyper-parameters by trial and error, including $L_j$, $F_j^{[l]}$, $I_j^{[l]}$, etc. We use the truncated Gaussian distribution as the differentiable pdf function, $\Psi_{\hat{\textbf{N}}}(\textbf{z}_1;\textbf{A}_1|\textbf{H},\boldsymbol{\lambda})$, $\Psi_{\hat{\textbf{M}}|\hat{\textbf{N}}}(\textbf{z}_2;\textbf{A}_2|\textbf{H},\textbf{z}_1)$, in both REGNNs to approximate $\textbf{N}$ and $\textbf{M}$. Then, the initial values of the trainable parameters of the cascaded REGNN $\textbf{A}_{j}^{\{0\}}, j\in \{1,2\}$ are randomly generated from a truncated Gaussian distribution with zero mean and 0.1 standard deviation.

\subsubsection{Generating the batch samples} We use $b$ batch samples to train the neural networks in each iteration. As the cascaded REGNN is trained in an unsupervised manner, the unlabeled training batch samples, including the average packet arrival rates $\boldsymbol{\lambda}^{\{ \tau \}}$ and the large-scale channel gains $\textbf{H}^{\{ \tau \}}$, are randomly generated in a simulation environment. 

\subsubsection{Computing the solutions} For each batch sample, given the realizations of $\boldsymbol{\lambda}^{\{ \tau \}}$ and $\textbf{H}^{\{ \tau \}}$, we can obtain the parameters of the distribution function of $N_k$, $(\xi_1^{\{ \tau \}}(k), \beta_1^{\{ \tau \}}(k)), k= 1,2,...,K$, from the first REGNN. Then, $\textbf{z}_1^{\{ \tau \}}$ can be generated according to the differentiable distribution. Given the input of the second REGNN, $\textbf{z}_1^{\{ \tau \}}$, we can obtain $(\xi_2^{\{ \tau \}}(k), \beta_2^{\{ \tau \}}(k))$ from its' output and generate $\textbf{z}_2^{\{ \tau \}}$ accordingly. To obtain the solutions to problem \eqref{op_in_training}, we convert the continuous outputs of the REGNNs to integers according to the following expressions, 
\begin{align}
\tilde{N}_{k}^{\{ \tau \}} & = \lceil {z}_{1,k}^{\{ \tau \}} N_k^{\rm q,\max} \rceil, \label{N_scale}\\
\tilde{M}_{k}^{\{ \tau \}} & = \lfloor {z}_{2,k}^{\{ \tau \}}\tilde{N}_{k}^{\{ \tau \}}\rceil,\label{M_scale}
\end{align}
where ${z}_{j,k}$ is the $k$th element of the vector $\textbf{z}_j$, $\lceil x \rceil$ is the smallest integer that is equal to or larger than $x$ and $\lfloor x \rceil$ is the nearest integer to $x$. 

\subsubsection{Probing the loss function} With the solution of each batch sample, we find the QoS violation probability for each link by three steps. Take $k$-th link as an example, we firstly find the queuing delay violation probability, $\epsilon_k^{\rm{q},\{ \tau \}}$, for a given input, $\lambda_k^{\{ \tau \}}$ and $\tilde{N}_k^{\{ \tau \}}$. Then, the policy is executed in the simulation for estimating the decoding error probability in \eqref{dep}. In each time slot, the $i$-th ($i\in K,i \neq k$) transmitter sends a packet with the probability of $\lambda_i^{\{ \tau \}} \tilde{M}_i^{\{ \tau \}}$. Finally, we count the number of transmissions that the sum of the decoding error probability and the queuing delay violation probability is larger than the reliability requirement of the $k$-th user, $\epsilon_k^{\max}$, to get the $P_k$ defined in \eqref{eq:P_k} and evaluate the QoS violation probability in \eqref{op_in_training}. The value of the loss function $f_{\rm {loss}}(.)$ is obtained from one sample by evaluating $\epsilon^{\rm q}_k$ and $\epsilon^{\rm c}_k$ in the objective function of problem \eqref{op_in_training}. 

\subsubsection{Updating the parameters} By estimating the gradients in \eqref{eq:gradientEstimation1} and \eqref{eq:gradientEstimation2}, we update the trainable parameters of the cascaded REGNN as follows,
\begin{align}\label{eq: graidnets update}
  \textbf{A}_{j}^{\{\tau+1\}} = \textbf{A}_{j}^{\{ \tau \}} + \zeta_j \nabla_{\textbf{A}_{j}}^{\{ \tau \}} \mathbb{E}_{\boldsymbol{\lambda},\textbf{H},\textbf{N},\textbf{M}} f_{\rm loss} \left ( \boldsymbol{\lambda}^{\{ \tau \}},\textbf{H}^{\{ \tau \}},\tilde{\textbf{N}}^{\{ \tau \}}, \tilde{\textbf{M}}^{\{ \tau \}} \right ),
\end{align}
where $\nabla_{\textbf{A}_{j}}^{\{ \tau \}}\mathbb{E}_{\boldsymbol{\lambda},\textbf{H},\textbf{N},\textbf{M}} f_{\rm loss} \left ( \boldsymbol{\lambda}^{\{ \tau \}},\textbf{H}^{\{ \tau \}},\tilde{\textbf{N}}^{\{ \tau \}}, \tilde{\textbf{M}}^{\{ \tau \}} \right )$ is obtained from \eqref{eq:gradientEstimation1} or \eqref{eq:gradientEstimation2}. The expectation in \eqref{eq: graidnets update} is estimated with a batch of samples.

\begin{algorithm}
 		\caption{ Unsupervised Cascaded REGNN Learning}\label{alg:A1}
        \begin{algorithmic}[1]
 		\INPUT $K$, $D_{\max}$, $T_s$, $W$, $a$, $P_k, k \in K$
 		\STATE Initialize: $L_j$, $F_j^{[l]}$, $I_j^{[l]}$, $\Psi_j$, $\textbf{A}_{j}^{\{0\}}$, $j\in{1,2}$, $l \in L$
 		\FOR{ $\tau = 1,2,3,...$ }
 		\STATE Generate the batch samples $\textbf{H}^{\{ \tau \}}$, $\boldsymbol{\lambda}^{\{ \tau \}}$.
 		\STATE Obtain the outputs $(\xi_j^{\{ \tau \}}(k), \beta_j^{\{ \tau \}}(k))$ and get the solutions $\tilde{\textbf{N}}^{\{ \tau \}}$ and $\tilde{\textbf{M}}^{\{ \tau \}}$.
 		\STATE Probe the loss function in \eqref{op_in_training}.
 		\STATE Update the parameters according to \eqref{eq: graidnets update}.
 		\ENDFOR
 		\OUTPUT $\textbf{A}_{j}, j\in \{1,2\}$
        \end{algorithmic}
\end{algorithm}

\section{Model-Based ES Method: A Stochastic Geometry Approach} 
\label{sec:analysis}
In this section, we derive two approximations of the QoS violation probability by using stochastic geometry, and apply them to simplify the objective function of problem \eqref{op}. Then, we use a model-based ES method to find the optimal solution. Note that the QoS violation probability in \eqref{op} does not have closed-form expression in general, we analyze the QoS violation probability in a symmetric scenario with the following three assumptions.

\begin{assumption}\label{as:ppp}
The distribution of transmitters follows a Poisson point process (PPP) $\Phi_x =\left\{{x_k}\in {{\mathbb R}^2}, k=0,1,2,... \right\}$ with intensity $\rho$.
\end{assumption}

\begin{assumption}\label{as:distance}
The distances between all the receivers and their corresponding transmitters are the same and are denoted by $r_0$.
\end{assumption}

Given the above two assumptions, the distribution of the receivers is also a PPP, $\Phi_y =\left\{ {y_k}\in {{\mathbb R}^2},k = 0,1,2,... \right\}$, with intensity $\rho$. Such a network model is referred to as Poisson bipolar network \cite[Definition~5.8]{stochastic2013Haenggi}.

\begin{assumption}\label{as:lambda}
The packet arrival processes at the transmitters are i.i.d. and are with the same average packet arrival rate $\lambda_0$.
\end{assumption}
The three assumptions indicate that the packet arrival processes, the large-scale channel gains, and the distributions of interference of all links are the same. Thus, the optimal transmission policies for different users should also be the same in this symmetric scenario. We denote the number of slots reserved for each packet and the number of repetitions of each packet by $N_0$ and $M_0$, respectively.

\subsection{Bounds of QoS Violation Probability}
To analyze the QoS violation probability, we consider a typical link where the typical transmitter is located at ${{x}_{0}}\in \Phi_x $ and the typical receiver is located at ${{y}_{0}}\in \Phi_y $. Let $\alpha$ be the path loss exponent, and $\mu_0={{r_0}^{-\alpha }}$ be the path loss of a typical link. The number of transmit antennas at the BS is one, $N_{\rm{T}}=1$. Let $g_0$ be the small-scale channel gain between $x_0$ and $y_0$, where $g_0=|h_{00}|^2$. We consider the Rayleigh fading channel, i.e., $g_0$, is an exponentially distributed random variable with parameter $1$. Without loss of generality, we normalize the transmit power to 1. Then, the receive power at the typical receiver $y_0$ is ${g_0}{{r_0}^{-\alpha }}$. The interference at the typical receiver is
\begin{equation}
{I}(t)=\sum\limits_{x_i\in {{\Phi }_{s}}\backslash \left\{ {{x}_{0}} \right\}}{{{g}_{i}}}{{r}_{i}}^{-\alpha }\mathbf{1}_i(t) ,
\end{equation}
where $\Phi_s\subset\Phi_x$ is the set of transmitters whose queues are non-empty in the considered time slot. 
$r_i$ is the distance between the transmitter $x_i$ and the typical receiver $y_0$. $g_i$ is the small-scale channel gain between the transmitter $x_i$ and the typical receiver $y_0$, where $g_i=|h_{i0}|^2$. In interference-limited networks, we ignore the thermal noise in the analysis. The SIR at the typical receiver $y_0$ is
\begin{equation}\label{eq:SIR}
\gamma_0(t)=\frac{{g_0}{r_0}^{-\alpha }}{\sum_{x_i\in {{\Phi }_{s}}\backslash \{ {{x}_{0}} \}}{{{g}_{i}}}{r_i}^{-\alpha }\mathbf{1}_i(t)} .
\end{equation}

According to the ergodicity of the PPP, the ensemble averages of the whole network equal the spatial averages of an arbitrary realization over a large region. Therefore, the probability that the typical link violates the QoS requirement equals the proportion of links in the network that cannot meet the QoS requirement, which is the definition of the QoS violation probability. The QoS violation probability is expressed as,
\begin{align}
{{P}_{\rm vio}}&={\mathbb{E}_{\Phi_s ,{{g}_{i}},{g_0}}} \mathbb{P} \left( \epsilon_0^{\rm q} + \epsilon_0^{\rm c} > \epsilon_0^{\max} \right)  \\ & ={\mathbb{E}_{\Phi_s ,{{g}_{i}},{g_0}}} \left[ \mathbb{P} \left( \epsilon_0^{\rm q} + \prod_{t=1}^{M_0}\epsilon_0^{\rm{d}}(t) > \epsilon_0^{\max} \right) \right] \label{eq:qos_analysis}\\
&={\mathbb{E}_{\Phi_s ,{{g}_{i}},{g_0}}} \left[ 1- \mathbb{P} \left(\epsilon_0^{\rm q}+ \prod_{t=1}^{M_0}\epsilon_0^{\rm{d}}(t) \le \epsilon_0^{\max}  \right) \right], 
\label{eq:qos_analysis_2}
\end{align}
where \eqref{eq:qos_analysis} is obtained from \eqref{dep}.

\begin{lemma}\label{lemma:1}
Given the number of time slots $N_0$ and the reliability threshold $\epsilon_0^{\max}$, if the decoding error probability of at least one of the $M_0$ repetitions, $\epsilon_0^{\rm d}(t)$, is less than $(\epsilon_0^{\max} - \epsilon_0^{\rm q})$,
then the reliability requirement can be satisfied, $\epsilon_0^{\rm q}+ \prod_{t=1}^{M_0}\epsilon_0^{\rm{d}}(t) \le \epsilon_0^{\max}$.
\begin{IEEEproof}
Since $\epsilon_0^{\rm{d}}(t) \leq 1, \forall t$, we have $\epsilon_0^{\rm q}+ \prod_{t=1}^{M_0}\epsilon_0^{\rm{d}}(t) \leq \epsilon_0^{\rm q} + \min_{t} \epsilon_0^{\rm{d}}(t)$.
If the decoding error probability of at least one of the $M_0$ repetitions is less than $(\epsilon_0^{\max} - \epsilon_0^{\rm q})$, then $\min_{t} \epsilon_0^{\rm{d}}(t) \leq \epsilon_0^{\max} - \epsilon_0^{\rm q}$. Therefore, we have $\epsilon_0^{\rm q}+ \prod_{t=1}^{M_0}\epsilon_0^{\rm{d}}(t) \leq \epsilon_0^{\rm q} + \min_{t} \epsilon_0^{\rm{d}}(t) \leq \epsilon_0^{\max}$. This completes the proof.
\end{IEEEproof}
\end{lemma}

Let's denote the event that the decoding error probability of at least one of the $M_0$ repetitions, $\epsilon_0^{\rm d}(t)$, is less than $(\epsilon_0^{\max} - \epsilon_0^{\rm q})$ by $\mathcal{A}_{\rm E}$, and denote the event that the reliability requirement can be satisfied by $\mathcal{B}_{\rm E}$. Lemma \ref{lemma:1} indicates that if $\mathcal{A}_{\rm E}$ happens, then $\mathcal{B}_{\rm E}$ will also happen. Thus, ${\mathbb{P}}(\mathcal{A}_{\rm E}) \leq {\mathbb{P}}(\mathcal{B}_{\rm E})$. Since the interference in different time slots is independent, we can obtain that ${\mathbb{P}}(\mathcal{A}_{\rm E}) = 1 - \left[\mathbb{P} \left( \epsilon_0^{\rm{d}}(t) > \epsilon_0^{\max} -\epsilon_0^{\rm q} \right)\right]^{M_0}$. Then, \eqref{eq:qos_analysis_2} can be expressed as
\begin{align}
  P_{\rm vio} & = {\mathbb{E}_{\Phi_s ,{{g}_{i}},{{g}_{0}}}}\{1- {\mathbb{P}}(\mathcal{B}_{\rm E})\} \nonumber\\
  & \leq {\mathbb{E}_{\Phi_s ,{{g}_{i}},{{g}_{0}}}}\{1- {\mathbb{P}}(\mathcal{A}_{\rm E})\}\nonumber\\
  &={\mathbb{E}_{\Phi_s ,{{g}_{i}},{{g}_{0}}}} \left[ 1- \mathbb{P} \left( \epsilon_0^{\rm{d}}(t) \le \epsilon_0^{\max} -\epsilon_0^{\rm q} \right) \right]^{M_0},  \nonumber\\ &={\mathbb{E}_{\Phi_s ,{{g}_{i}},{{g}_{0}}}} \left[ 1-\mathbb{P} \left(  \epsilon_0^{\rm{d}} \le \epsilon_{\rm th}^{\rm d} \right) \right]^{M_0}\triangleq P^{\rm up}_{\rm vio},\label{eq:pvio}
\end{align}
where $P^{\rm up}_{\rm vio}$ is a upper bound of $P_{\rm vio}$, the time index in $\epsilon_0^{\rm{d}}(t)$ is ignored since it is i.i.d. in different slots, and $\epsilon_{\rm th}^{\rm d}$ is the required threshold of the decoding error probability of a single transmission,
\begin{equation}\label{eq:dep_th}
\epsilon_{\rm th}^{\rm d}\triangleq ~{\epsilon_0^{\max} - \epsilon_0^{\rm{q}}}.
\end{equation}
Here we define a SIR threshold, $\gamma_{\rm th}$, which is the minimum SIR that satisfies $\epsilon_0^{\rm{d}} \le \epsilon_{\rm th}^{\rm d}$. By analyzing the probability that the SIR is lower than the threshold, we can obtain the expectation in \eqref{eq:pvio}.
The SIR threshold is decided by $N_0,~ \epsilon_0^{\rm{q}}, ~\epsilon_{\rm th}^{\rm{d}}$. Given $N_0$ and the queuing delay bound $D_{k,\max}^{\rm{q}}$ we could find the queuing violation probability $ \epsilon_0^{\rm{q}}$ following the relationship in \eqref{eq:dvp_con} by a binary search method. With the value of the queuing violation probability, we could get the required threshold of the decoding error probability of a single transmission $\epsilon_{\rm th}^{\rm{d}}$ via \eqref{eq:dep_th}. Setting $\epsilon_{0}^{\rm d}(t)=\epsilon_{\rm th}^{\rm{d}}$, the achievable rate $R_0$ is calculated via \eqref{R} under different values of SIR, $\gamma_{0}$. The SIR monotonically increases when $R_0$ increases. Thus, with the binary search method, we could find the SIR threshold $\gamma_{\rm th}$ by solving the following problem, 
\begin{align}\label{eq:threshold}
\gamma_{\rm th} = & \min_{\gamma_{0}}~  \gamma_{0}\\
s.t. ~& R_0(\gamma_{0},\! \epsilon_{\rm th}^{\rm d}) \ge 1 \tag{\theequation a} \\
& \eqref{R},\eqref{eq:dvp_con},\eqref{eq:dep_th}. \tag{\theequation b} 
\end{align} 
Given the SIR threshold, the upper bound of the QoS violation probability can be re-expressed as follows,
\begin{align}
{{P}_{\rm vio}^{\rm up}}&={\mathbb{E}_{\Phi_s ,{{g}_{i}},{{g}_{0}}}}{{\left[ 1-\mathbb{P} \left( \text{SIR} \ge \gamma_{\rm th} \right) \right]}^{M_0 }}\\
&\overset{(a)}{=}{\mathbb{E}_{\Phi_s ,{{g}_{i}},{{g}_{0}}}}{{\left[ 1-\mathbb{P} \left( {{g}_{0}}\ge \gamma_{\rm th}{r_0}^{\alpha }I\left| \Phi_s ,{{g}_{i}},{{g}_{0}} \right. \right) \right]}^{M_0 }},
\label{ThmTDRP}
\end{align}
where (a) is obtained from \eqref{eq:SIR}. Since \eqref{ThmTDRP} does not have a closed-form expression, we turn to derive two lower bounds of ${{P}_{\rm vio}^{\rm up}}$, as the two approximations of ${P}_{\rm vio}$.

\begin{theorem}
\label{thm:lower1}
Approximation $1$ of $P_{\rm vio}$ is
\begin{equation}
 {{P}_{\rm A1}}={{\left( 1-\exp \left( -\frac{M_0}{N_0} C \right) \right)}^{M_0 }},
\label{eq:A1}
\end{equation}
where $C=\pi \rho \lambda N_0 {r_0^2}{{\gamma_{\rm th}}^{\frac{2}{\alpha }}}\Gamma (1+\frac{2}{\alpha })\Gamma (1-\frac{2}{\alpha })$. $\Gamma(x)$ is the standard gamma function, $\Gamma(x)=\int_0^{\infty}t^{x-1}e^{-t}dt$.
\end{theorem}
\begin{IEEEproof}
According to the Jensen's inequality, if $x$ is a random variable and $\varphi$ is a convex function, we have $\varphi \left( \mathbb{E}\left[ x \right] \right)\le \mathbb{E}\left[ \varphi \left( x \right) \right]$. Since $f\left( x \right)={{\left( 1-x \right)}^{M_0 }}$ is a convex function, we have $ f\left( \mathbb{E}\left[ x \right] \right)\le \mathbb{E}\left[ f\left( x \right) \right]$, where
\begin{equation}
x=\mathbb{P}\left( {{g}_{0}}\ge \gamma_{\rm th}{r_0}I\left| {{\Phi }_{s}},{{g}_{i}},{{g}_{0}} \right. \right).
\end{equation}
Thus, we can obtain a lower bound of \eqref{ThmTDRP} which is the approximation of $P_{\rm vio}$ as follows,
\begin{equation}
\begin{aligned}
P_{\rm vio}^{\rm up} & ={{E}_{{{\Phi }_{s}},{{g}_{i}},{{g}_{0}}}}{{\left( 1-\mathbb{P}\left( {{g}_{0}}\ge \gamma_{\rm th}{r_0}^{\alpha }I\left| {{\Phi }_{s}},{{g}_{i}},{{g}_{0}} \right. \right) \right)}^{M_0 }} \\
 & \ge {{\left( 1-{{E}_{{{\Phi }_{s}},{{g}_{i}},{{g}_{0}}}}\mathbb{P}\left( {{g}_{0}}\ge \gamma_{\rm th}{r_0}^{\alpha }I\left| {{\Phi }_{s}},{{g}_{i}},{{g}_{0}} \right. \right) \right)}^{M_0 }} \\
 & ={{\!\Big(\! 1\!-\!{{E}_{{{\Phi }_{s}}}}\prod\limits_{x\in {{\Phi }_{s}}\backslash \!\left\{\! {{x}_{0}} \!\right\}\!}{\!\Big(\! \frac{M_0}{N_0} \frac{1}{1\!+\!\gamma_{\rm th}{r_0}^{\alpha }{r_i}^{-\alpha }}+1-\frac{M_0}{N_0}  \!\Big)\!} \!\Big)\!}^{M_0 }} \\
 & =\!{{\!\left(\! 1\!-\!\exp \!\left(\! \!-\!\rho \int_{{{R}^{2}}}{1\!-\!\!\left(\! \frac{M_0}{N_0} \frac{1}{1\!+\!\gamma_{\rm th}{r_0}^{\alpha }{r_i}^{\!-\!\alpha }}\!+\!1\!-\!\frac{M_0}{N_0}  \!\right)\!dr} \right) \right)}^{M_0 }} \\
 & ={{\left( 1-\exp \left( -\frac{M_0}{N_0} C \right) \right)}^{M_0 }},
\end{aligned}
\label{}
\end{equation}
which is the approximation in Theorem \ref{thm:lower1}.
\end{IEEEproof}

\begin{theorem}\label{thm:lower2}
Approximation $2$ of $P_{\rm vio}$ is obtained under the assumption that only the nearest active transmitter generates interference to the typical receiver,
\begin{equation}
{{P}_{\rm A2}}=\!Z\!-\!Z\int_{0}^{\!+\!\infty }{\frac{{{f}^{\rm t}_{1}}\left( r_1 \right)}{1\!+\!\gamma_{\rm th} r_{0}^{\alpha }r_{1}^{\!-\!\alpha }}} d{{r}_{1}},
\label{eq:A2}
\end{equation}
where $r_1$ is the distance between the typical receiver and the nearest active transmitter, ${{f}^{\rm t}_{1}}\left( r_1 \right) =2\pi \rho \lambda N_0 {r_1}\exp \left( -\rho \lambda N_0 \pi r_1^{2} \right)$ is the distribution of $r_1$,
and $Z$ is the probability that the typical transmitter and the nearest active transmitter choose the same slots to repeat their packets. Since both transmitters select $M_0$ slots from $N_0$ slots, the probability that the selected slots are the same can be expressed as $Z={N_0\choose M_0}^{-1}$, where ${N_0\choose M_0}$ is the number of combinations when selecting $M_0$ slots from $N_0$ slots.
\end{theorem}
\begin{IEEEproof}
In order to derive Approximation $2$, we only consider the interference from the nearest active transmitter, and the interference from other links is ignored. Since the interference is less than that in the original network model, we can obtain a lower bound of \eqref{ThmTDRP} and use it as the approximation of the QoS violation probability. Thus, the approximation can be expressed as follows,
\begin{align}
  {{P}_{\rm A2}}
  & =(1-Z)\times \mathbb{P}_{\rm vio}^{'} +Z\times \left(1-\mathbb{P}\left( {{g}_{0}}\ge  \gamma_{\rm th} {r_0^{\alpha }}{{g}_{1}}r_{1}^{-\alpha } \right) \right)\label{eq:PA2twoterm}\\ & \geq  \! Z\!\times\left(1-\mathbb{P}\left( {{g}_{0}}\ge \gamma_{\rm th} {r_0^{\alpha }}{{g}_{1}}r_{1}^{-\alpha } \right)\right), \label{eq:PA2oneterm}
\end{align}
where ${g}_{1}$ is the small-scale channel gain between the typical receiver and the nearest active transmitter, and $\mathbb{P}_{\rm vio}^{'}$ is probability that the SIR is lower than $\gamma_{\rm th}$ in a time slot with no interference. When the typical transmitter and the nearest active transmitter select different time slots, there is no interference in at least one slot. Thus, the first term in \eqref{eq:PA2twoterm} is the QoS violation probability in this case. The second term in \eqref{eq:PA2twoterm} is the QoS violation probability in the other case that both transmitters select the same slots to repeat their packets. Since $\mathbb{P}_{\rm vio}^{'}$ is small, we can obtain \eqref{eq:PA2oneterm} by ignoring the first term. Then, we obtain the approximation of \eqref{eq:PA2twoterm} as follows,
\begin{equation}
    \begin{aligned}
 {{P}_{\rm A2}} & = Z \times \left( 1- \mathbb{E}_{r_1} \left[ \mathbb{P}\left( {{g}_{0}}\ge \gamma_{\rm th} {r_0^{\alpha }}{{g}_{1}}r_{1}^{-\alpha } |r_1 \right) \right] \right) \\
  &= Z \times \left( 1 -\int_{0}^{\!+\!\infty }{{f}^{\rm t}_{1}}(r_1) \mathbb{E}_{g_0,g_1} \left[ \mathbb{P}\left( {{g}_{0}}\ge \gamma_{\rm th} {r_0^{\alpha }}{{g}_{1}}r_{1}^{-\alpha } |g_0,g_1 \right) \right]d{{r}_{1}} \right) \\
  &= Z \times \left( 1 -\int_{0}^{\!+\!\infty }{{f}^{\rm t}_{1}}(r_1)  \mathbb{E}_{g_1}\left[ {\rm exp} \left(-\gamma_{\rm th} {r_0^{\alpha }}r_{1}^{-\alpha } g_1 \right)\right] d{{r}_{1}} \right) \\
 & =\!Z\!-\!Z\!\int_{0}^{\!+\!\infty }{{f}^{\rm t}_{1}}\left( r_1 \right)\frac{1}{1\!+\!\gamma_{\rm th} r_{0}^{\alpha } r_{1}^{\!-\!\alpha }} d{{r}_{1}}\label{eq:PA2integral},
 \end{aligned}
\end{equation}
Thus, we obtain the approximation in Theorem \ref{thm:lower2}.
\end{IEEEproof}

\subsection{Problem Formulation in the Symmetric Scenario}
Considering that the QoS violation probability, $P_{\rm vio}$ in \eqref{op}, does not have a closed-form expression, we turn to minimize $\max\{P_{\rm A1},P_{\rm A2}\}$. Given the two approximations, the optimization problem in Section \ref{sec:problem formulation} can be re-expressed as follows,
\begin{align}\label{op_analysis}
\min_{N_0,M_0} 
  &~~~
  \max \{{P}_{\rm A1},{P}_{\rm A2}\}\\
 \text{s.t} ~~
  & 1 \leq N_0 \leq N_0^{\rm q,\max}, k=1,2,...,K,  \tag{\theequation a}\\
  & 1\leq M_0 \leq N_0, k=1,2,...,K, \tag{\theequation b}\\
  & \eqref{eq:dvp_con},~\eqref{eq:dep_th}.\nonumber
\end{align}	
Although the problem is still a non-linear integer programming, the feasible region is a two-dimensional space, and the objective function can be obtained numerically using the approximations. Thus, we can use ES to solve it.

\section{Simulations}
In this section, we first introduce two types of network topologies: a random network model and a hexagonal cellular network model. Then, we present the training setup of the cascaded REGNN and the training results. Finally, we evaluate the performance of cascaded REGNN and ES methods in different scenarios. Specifically, we test the cascaded REGNN and ES method in both the symmetric scenarios and more general scenarios without the three assumptions in Section IV.

\subsection{Simulation Setup}
We first set up the system parameters. The duration of a time slot is $T_{\rm s} = 0.1$~ms. The E2E delay bound is set to be $D_{\max} = 5$~ms. There are 128 bits in one packet. There are $N_{\rm T}=16$ antennas at the transmitter side. The transmit power of each transmitter is $p_k=23$~dBm. The transmission bandwidth for each link is $1$~MHz. The cell density is $\rho = 28~ \text{BSs}/{\text{km}^2}$. The average packet arrival rate, $\lambda_k$, is uniformly distributed between $0.01$ and $0.1$~packets/slot. 

For the first topology, the locations of transmitters are randomly generated in a fixed area with density $\rho$. This is referred to as \textbf{Topology I}. Let $\alpha$ be the path loss exponent, and $\mu_0={r_{ki}^{-\alpha }}$. $r_{ki}$ (m) is the distance between the transmitter of $k$-th link and the receiver of the $i$-th link. The path loss exponent is $\alpha=4$.

\begin{figure}[h]
\centering
\includegraphics[width=0.5\textwidth]{ 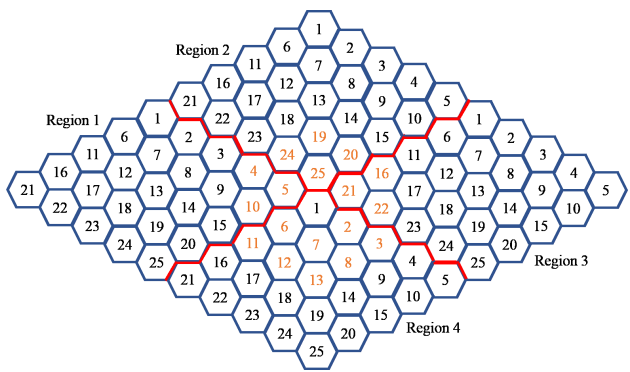}
\caption{The network topology of hexagonal cellular network. Region 1 in Fig. \ref{fig:hexcell} is a minimum unit of the cellular network where there are 25 links (BS-user pairs). This figure illustrates how to fill up the whole area with the minimum unit. We label the cells from 1 to 25. Cell $i,\forall i=1,2,...,25$ in different regions use the same repetition scheme. In the simulation, we only consider strong interference from nearby cells. For example, cell 1 in region 4 only suffers interference from the cells in the orange color.}\label{fig:hexcell}
\end{figure}

The second topology is the hexagonal cellular network referred to as \textbf{Topology II}, which is shown in Fig.~\ref{fig:hexcell}. The radius of the cell is $100$~m. The path loss model is $35.3+37.6\log_{10}{r_{ki}}$. 

In both Topology~I and II, we randomly generate the receiver's location in each cell. The distances between the receivers and the transmitters are random values between $50$~m and $100$~m. For the small-scale channel gain, we use the Nakagami-$m$ fading channel with $m=3$. The distribution of $\textbf{h}_{i,j}^*\textbf{h}_{i,j}, i,j=1,...,K$ can be expressed as \cite{goldsmith2005wireless}
\begin{align}
    f_{\textbf{h}_{i,j}}(x) = \frac{m^{N_{\rm T}m}x^{N_{\rm T}m-1}}{(N_{\rm T}m-1) !\ }\exp(-mx)\label{eq:channelgain}.
\end{align} 

\subsection{Training of Cascaded REGNN}
In this subsection, we train the cascaded REGNN in two different topologies. The training process follows the unsupervised learning algorithm in Algorithm \ref{alg:A1}. We choose the truncated Gaussian distribution to replace the deterministic policies $\textbf{N}(\textbf{H}, \boldsymbol{\lambda})$ and $\textbf{M}(\textbf{H}, \textbf{N})$. The truncated Gaussian distribution's lower and upper bound are $0$ and $1$. The activation function in the output layer is a sigmoid function, $\psi_j^{[L]}(x) = \frac{1}{1+e^{-x}}$. The $k$-th output of $j$-th REGNN is the mean value of the truncated Gaussian distribution for the $k$-th user, $\xi_j(k)$. The $(K+k)$-th output of $j$-th REGNN is the standard deviation of the distribution functions for the $k$-th user, $\beta_j(k)$. The number of batch samples in each iteration is $b=64$. The hyper-parameters of the cascaded REGNN are listed in Table \ref{training_parameters}. As the input and output dimensions of the two functions $\textbf{N}$ and $\textbf{M}$ are the same, we use the same structure of the convolutional filter in the two REGNNs, including the number of hidden layers, the number of features in them, and filter length. Through trial and error, we choose the best combination of the hyper-parameters for the cascaded REGNNs. There are $\sum_{j=1}^2 \sum_{l=1}^{25}F_j^{[l]}F_j^{[l+1]}I_j^{[l]}=2\times(1\times2\times4+\sum_{l=2}^{25} 2\times2\times4)=784$ trainable parameters, which do not increase with the number of wireless links in the network. After the training stage, the cascaded REGNN is then used to generate $\textbf{N}$ and $\textbf{M}$ according to the realizations of $\boldsymbol{\lambda}$ and $\textbf{H}$. 

\begin{table}[h]
	\centering
	\caption{ Hyper-parameters of the cascaded REGNN}
	\label{training_parameters}
	\begin{tabular}{|c|c|c|}
		\hline
		Parameter & REGNN1 & REGNN2\\ \hline
		Number of layers $(L_j)$ &25 &25\\ \hline
		Filter length $(I_j^{[l]})$ &4 & 4 \\ \hline
		Number of features $(F_j^{[l]})$ & 2 & 2\\ \hline
		Learning rate ($\zeta_j$) & 0.001 & 0.001\\ \hline
		Active function in hidden layers ($\psi_j^{[l]}$) & Relu & Relu \\ \hline
	\end{tabular}
\end{table}

\begin{figure}[h]
	\centering
	\includegraphics[width=0.6\textwidth]{ 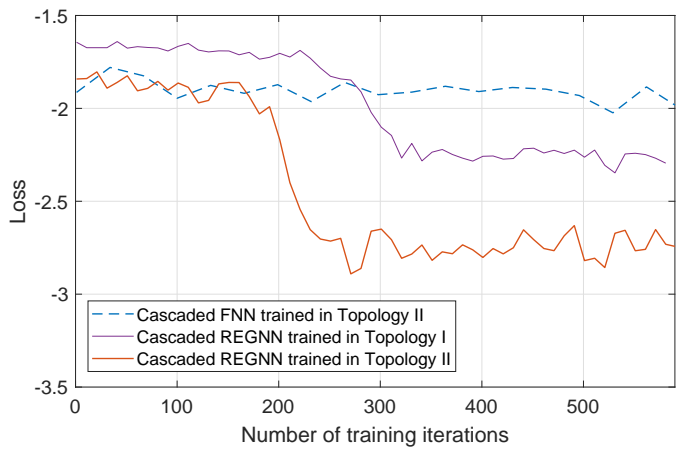}
	\caption{Performance of the loss in the training stage.}\label{Fig:Cost_Fun}
\end{figure}

The values of the loss function during the training process are shown in Fig. \ref{Fig:Cost_Fun}. We compare the convergence time of the cascaded REGNN with a cascaded FNN structure, where the REGNNs in Fig. \ref{Fig:C_GNN} are replaced with two FNNs. As shown in Fig. \ref{Fig:Cost_Fun}, the loss function of the cascaded REGNN trained in Topology II converges faster than the cascaded REGNN trained in Topology I. Thus, we train the cascaded FNN in Topology II and compare its performance with that of the GNN. The input of the first FNN is a $650 \times 1$ vector with large-scale channel gains and arrival rates $\boldsymbol{\lambda}$. The input of the second FNN is also a $650 \times 1$ vector with large-scale channel gains and $\textbf{z}_1^{\{ \tau \}}$. The outputs of the two cascaded FNN are the same as the cascaded REGNN, which are the parameters of the distribution function of $N_k$, $(\xi_1^{\{ \tau \}}(k), \beta_1^{\{ \tau \}}(k)), k= 1,2,...,K$ and the parameters of the distribution function of $M_k$, $(\xi_2^{\{ \tau \}}(k), \beta_2^{\{ \tau \}}(k))$ in the $\tau$-th training iteration. The hyper-parameters of the two FNNs are the same. For each FNN, there are three fully connected hidden layers with the sizes of $256$, $128$ and $64$, respectively. The active function is Relu and the learning rate is the same as that of the cascaded REGNN in Table \ref{training_parameters}. The total number of trainable parameters is about $400,000$.

The results in Fig. \ref{Fig:Cost_Fun} show that the loss functions of the cascaded REGNN trained in Topology~I and Topology~II converge to about $-2.25$ and $-2.8$ after $370$ and $270$ iterations, respectively. However, the loss function with cascaded FNNs decreases slowly and does not converge after $500$ training iterations. This is because the number of trainable parameters in cascaded FNN is much larger than that in cascaded REGNN, and the network's topology is not exploited to reduce the parameters.

\subsection{Performance Evaluation in Symmetric Scenarios}\label{sec:numerical_eva}
In this subsection, we evaluate the QoS violation probabilities obtained by the cascaded REGNN and the ES method in the symmetric scenarios, which are the special cases of Topology I. The three assumptions in Section~\ref{sec:analysis} are needed to apply the ES method: 1) the distribution of the transmitters follows a PPP with intensity $\rho=28~ \text{BSs}/{\text{km}^2}$; 2) the distance between each transmitter and its receiver is fixed at $75$~m; 3) the average packet arrival rate for all the transmitters are the same as $\lambda_0=0.05$~packets/slot.

\begin{figure}[h]
	\centering
	\includegraphics[width=0.6\textwidth]{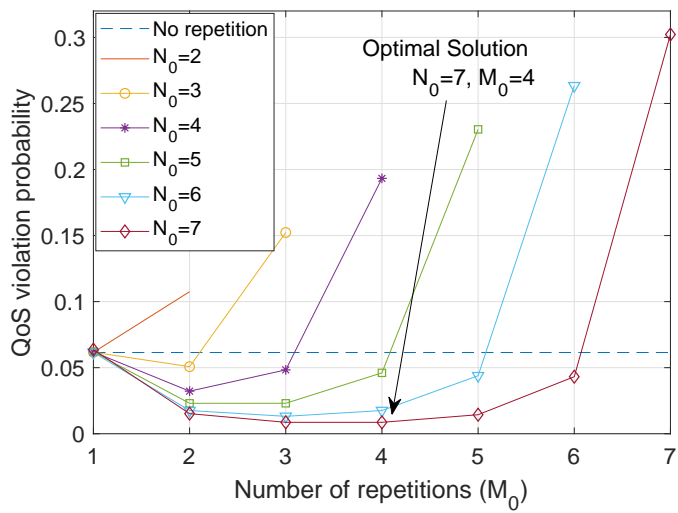}
	\caption{The QoS violation probability as functions of the number of repetitions $M_0$. 
	}\label{fig:ex_a}
\end{figure}

We use ES to find the solution to the optimization problem \eqref{op_analysis}. Figure \ref{fig:ex_a} shows the QoS violation probability with different values of $N_0$ and $M_0$. The dotted blue line is the QoS violation probability achieved by the no-repetition scheme, i.e., $N_0=M_0=1$. When $N_0=M_0>1$, the repetition scheme degenerates to the K-repetition scheme. With the K-repetition scheme, the QoS violation probability increases with repetitions. This is because the queuing delay violation probability increases with the number of repetitions, but the packet loss probability does not decrease when there is no diversity gain. With the optimal random repetition scheme $N_0=7$ and $M_0=4$ the QoS violation probability is $0.008634$. The performance gain is around $86\%$ when compared with the no repetition scheme ($P_{\rm vio}=0.0615$). 

In Table \ref{Table: Density Performance Comparison}, we train the cascaded REGNN in a symmetric network with the cell density of $\rho = 28~\text{BSs}/\text{km}^2$. Then, we test the cascaded REGNN in symmetric networks with different cell densities. We also evaluate the QoS violation probabilities achieved by the optimal policy obtained from the model-based ES method.
The QoS violation probabilities are evaluated with $10^3$ realizations of interference networks. The mean QoS violation probabilities of the $10^3$ realizations are shown in Table \ref{Table: Density Performance Comparison}. The results indicate that the QoS violation probabilities achieved by the two methods are nearly the same. Even if we change the cell density in the testing, the cascaded REGNN is near-optimal. The results imply that the cascaded REGNN generalizes well to wireless networks with different cell densities.

\begin{table}[h]
 	\centering
	\caption{QoS violation probabilities with different cell densities: Cascaded REGNN vs benchmarks }
 	\label{Table: Density Performance Comparison}
 	\begin{tabular}{|c|c|c|c|c|}		
 		 \hline
 		 Densities of BSs in testing & Cascaded REGNN & ES ($N_0,M_0$)\\ \hline
 		  $\rho = 14~\text{BSs}/\text{km}^2$ & 0.0016 & 0.0016 (7,4) \\ \hline 
 		 $\rho = 28~\text{BSs}/\text{km}^2$ & 0.0057 & 0.0056 (7,4)  \\ \hline
         $\rho = 56~\text{BSs}/\text{km}^2$ &  0.0316 & 0.0309 (7,5) \\ \hline
	\end{tabular}
\end{table}

\subsection{Performance Evaluation in General Scenarios}
In this subsection, we evaluate the QoS violation probabilities achieved by the cascaded REGNN and the ES method in more general scenarios (without the three assumptions in Section IV). To validate the generalization ability of the cascaded REGNN, the network scales, frequency reuse factors, or the network topologies in the testing stage are different from that in the training stage. To illustrate the impact of model mismatch on the performance of the model-based ES method, we obtain the solution with the help of the three assumptions in Section IV and evaluate the QoS violation probabilities in networks without these assumptions.

In Table \ref{Table: Scale Performance Comparison}, we train the cascaded REGNN in Topology II with 25 wireless links and test it in Topology II with more links. The topology of the network with $100$ links is shown in Fig \ref{fig:hexcell}. To obtain the results with $50$ links, we only evaluate the QoS violation probability in Region 1 and Region 2. It is worth noting that the hyper-parameters in Table \ref{training_parameters} and all the parameters in the cascaded REGNN are fixed in the testing stage. The density of the network remains the same. We compare the QoS violation probability achieved by the cascaded REGNN with that achieved by the no-repetition scheme (i.e., $N_k=1, M_k=1,\forall k$) and the K-repetition scheme (i.e., $N_k=5$ and $M_k=5$). For the K-repetition scheme, we set $M_k=N_k$ and select the maximum $N_k$ that satisfis the delay requirement. For a fair comparison, the E2E delay bound is the same for the three policies. When $K=25$, the performance gain in terms of the mean QoS violation probabilities is close to $74$\% compared with the no repetition scheme, and the performance gain is around $20$\% compared with the K-repetition scheme. The results in Table \ref{Table: Scale Performance Comparison} also indicate that the QoS violation probability only changes slightly with the number of links. This validates that the cascaded REGNN can be applied in wireless networks with different numbers of links without fine-tuning.

\begin{table}[h]
	\centering
	\caption{Performance comparison of the QoS violation probability : Cascaded REGNN vs Existing Policies}
	\label{Table: Scale Performance Comparison}
	\begin{tabular}{|c|c|c|c|c|c|}
		\hline
	    Number of links  & Cascaded REGNN trained in Topology II & No repetition & Gain & K-repetition & Gain\\ \hline
		$K=25$ & 0.0072 & 0.0274& 73.72\% & 0.0090 & 20.0\%\\ \hline 
		$K=50$ &  0.0069 & 0.0272& 74.63\% & 0.0086 & 19.77\% \\ \hline	
		$K=100$ &  0.0071 & 0.0276 & 74.28\% & 0.0088 & 19.32\% \\ \hline 
	\end{tabular}
\end{table}

Furthermore, our learning method can be applied to networks with different frequency-reuse factors. It is worth noting that even if the frequency-reuse factor is not $1$, inter-cell interference should be considered in ultra-dense networks. Simulation results under different frequency reuse factors ($1$, $1/3$, $1/7$) in Topology~II are provided in Table~\ref{freuency_reuse}, where the number of links is $25$. As shown in Table~\ref{freuency_reuse}, the gain of the cascaded REGNN compared with the K-repetition only decreases slightly from $20\%$ to $18\%$ when we change the frequency reuse factor from $1$ to $1/3$. When the frequency reuse factor is $1/7$, the QoS violation probabilities achieved by both schemes are much higher than the previous scenarios. This is because the bandwidth of each cell decreases as the frequency reuse factor decreases. If the bandwidth of a cell is too small, the decoding error probabilities will be much higher than the requirement of URLLC. 
\begin{table}[h]
	\centering
	\caption{Performance comparison of the QoS violation probability: Cascaded REGNN vs Existing Policies}
	\label{freuency_reuse}
	\begin{tabular}{|c|c|c|c|c|c|c|c|c|}
		\hline
	    frequency reuse factor & cascaded REGNN trained in Topology II   & K-repetition & Gain\\ \hline
	    $1$ & 0.0072  & 0.0090 & 20\% \\ \hline
	    $1/3$ & 0.000702   & 0.000865 &  18.4\%\\ \hline
	    $1/7$ & 0.2833  &  0.3246 & 12.72\%\\ \hline
	\end{tabular}
\end{table}

\begin{figure}[h]
	\centering
	\includegraphics[width=0.6\textwidth]{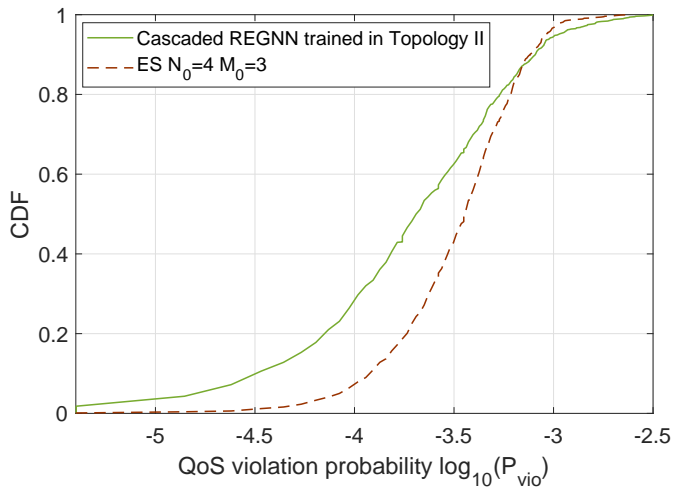}
	\caption{CDF of the QoS violation probability.
	} \label{Fig:model_mismatch}
\end{figure}

To illustrate the impacts of model mismatch on the QoS violation probability, we train the cascaded REGNN in Topology II and test it in Topology I. Specifically, we train the cascaded REGNN in Topology II with $25$ wireless links and $28~\text{BSs}/\text{km}^2$ cell density. The network for testing has $K=100$ links, and $\rho=56~\text{BSs}/\text{km}^2$. For comparison, we find the optimal solution, $N_0$ and $M_0$, by the ES method in a symmetric scenario. We set $r_0=75$~m, $\lambda_0=0.1$~packet/slots, and $\rho=56~\text{BSs}/\text{km}^2$. The ES solution is also tested in Topology I with 
$K=100$ and $\rho=56~\text{BSs}/\text{km}^2$.
The cumulative distribution functions (CDFs) of the QoS violation probability obtained from $10^3$ realizations are shown in Fig. \ref{Fig:model_mismatch}. 
The mean values of the QoS violation probabilities achieved by the cascaded REGNN trained in Topology II and ES ($N_0=4,~M_0=3$) are $3.52\times10^{-4}$ and $4.16\times10^{-4}$, respectively. The results indicate that the cascaded REGNN trained in Topology II outperforms the ES by $15\%$. Thus, we can conclude that the cascaded REGNN is more robust to the model mismatch than the model-based ES method.

\section{Conclusion}
In this paper, we minimized the QoS violation probability by optimizing a random repetition policy. To solve the optimization problem, we used a cascaded REGNN to represent the repetition policy in wireless networks. We trained the cascaded REGNN via a model-free unsupervised learning algorithm. We applied stochastic geometry to derive the approximations of the QoS violation probability in a symmetric Poisson bipolar network. ES was implemented to find the optimal repetition policy. Our simulation results showed that the cascaded REGNN achieves nearly the same performance as the ES method in symmetric networks with different cell densities. In more general networks, the cascaded REGNN reduces the QoS violation probability by $74$\% and $20$\% compared with the scheme without repetition and the K-repetition scheme, respectively. Furthermore, our results indicated that the cascaded REGNN generalizes very well to wireless networks with different scales, network topologies, cell densities, and frequency reuse factors. It outperforms the model-based ES method in the presence of the model mismatch.

\appendices
\renewcommand{\theequation}{A.\arabic{equation}}
\setcounter{equation}{0}
\section*{Appendix A: Derivations of \eqref{eq:gradientEstimation1} and \eqref{eq:gradientEstimation2}}
\label{appen_1}
By replacing the deterministic policies with the stochastic policies, the realizations of $\textbf{N}$ and $\textbf{M}$ are denoted by $\textbf{z}_1$ and $\textbf{z}_2$, respectively.
The expectation in \eqref{eq:gradientEstimation1} can be expressed as follow, 
\begin{align} 
    &\nabla_{\textbf{A}_{1}} \mathbb{E}_{\boldsymbol{\lambda},\textbf{H},\textbf{N},\textbf{M}} f_{\rm loss} \left ( \boldsymbol{\lambda},\textbf{H},\textbf{N}, \textbf{M} \right )  \nonumber\\
    =&\nabla_{\textbf{A}_{1}} \mathbb{E}_{\boldsymbol{\lambda},\textbf{H}} \int_{\textbf{z}_1, \textbf{z}_2} f_{\rm loss} \left ( \boldsymbol{\lambda},\textbf{H},\textbf{z}_1,\textbf{z}_2\right ) \! \Psi_{\hat{\textbf{N}},\hat{\textbf{M}}} ~  d{\textbf{z}_1}  d{\textbf{z}_2},  \label{eq:variable_transforming_2}
\end{align}
where $\Psi_{\hat{\textbf{N}},\hat{\textbf{M}}}$ is the joint probability density functions of the stochastic policies $\hat{\textbf{N}}$ and $\hat{\textbf{M}}$. The joint pdf could be calculated as
\begin{align}
    \Psi_{\hat{\textbf{N}},\hat{\textbf{M}}}=\Psi_{\hat{\textbf{N}}}(\textbf{z}_1;\textbf{A}_1|\textbf{H},\boldsymbol{\lambda})\Psi_{\hat{\textbf{M}}|\hat{\textbf{N}}}(\textbf{z}_2;\textbf{A}_2|\textbf{H},\textbf{z}_1).\label{eq:condpdf}
\end{align}
By substituting \eqref{eq:condpdf} to \eqref{eq:variable_transforming_2} and changing the order of the integrals and the derivative, we have
\begin{align}
    &\nabla_{\textbf{A}_{1}} \mathbb{E}_{\boldsymbol{\lambda},\textbf{H},\textbf{N},\textbf{M}} f_{\rm loss} \left ( \boldsymbol{\lambda},\textbf{H},\textbf{N}, \textbf{M} \right)  \nonumber \\
    =& \mathbb{E}_{\boldsymbol{\lambda},\textbf{H}} \int_{\textbf{z}_1,\textbf{z}_2} f_{\rm loss} \left ( \boldsymbol{\lambda},\textbf{H},\textbf{z}_1,\textbf{z}_2\right ) \nabla_{\textbf{A}_{1}} [\Psi_{\hat{\textbf{N}}}(\textbf{z}_1;\textbf{A}_1|\textbf{H},\boldsymbol{\lambda})
    ] \Psi_{\hat{\textbf{M}}|\hat{\textbf{N}}}(\textbf{z}_2;\textbf{A}_2|\textbf{H},\textbf{z}_1) ~ d{\textbf{z}_1}  d{\textbf{z}_2}.
\end{align}
Then we only need to calculate the derivative of $\Psi_{\hat{\textbf{N}}}(\textbf{z}_1;\textbf{A}_1|\textbf{H},\boldsymbol{\lambda})$ with respect to $\textbf{A}_1$. By using the property that $\nabla_{x}f(x) = f(x)\nabla_{x} \log f(x) $, we can derive that
\begin{align}
&\nabla_{\textbf{A}_{1}} \mathbb{E}_{\boldsymbol{\lambda},\textbf{H},\textbf{N},\textbf{M}} f_{\rm loss} \left ( \boldsymbol{\lambda},\textbf{H},\textbf{N}, \textbf{M} \right )  \nonumber \\
 =& \mathbb{E}_{\boldsymbol{\lambda},\textbf{H}} \int_{\textbf{z}_1,\textbf{z}_2} f_{\rm loss} \left ( \boldsymbol{\lambda},\textbf{H},\textbf{z}_1,\textbf{z}_2\right ) \Psi_{\hat{\textbf{N}}}(\textbf{z}_1;\textbf{A}_1|\textbf{H},\boldsymbol{\lambda})\Psi_{\hat{\textbf{M}}|\hat{\textbf{N}}}(\textbf{z}_2;\textbf{A}_2|\textbf{H},\textbf{z}_1) \nabla_{\textbf{A}_{1}} [\Psi_{\hat{\textbf{N}}}(\textbf{z}_1;\textbf{A}_1|\textbf{H},\boldsymbol{\lambda})
    ] ~ d{\textbf{z}_1}  d{\textbf{z}_2} \nonumber \\
=& \mathbb{E}_{\boldsymbol{\lambda},\textbf{H},\textbf{z}_1,\textbf{z}_2} \left\{f_{\rm loss} \left ( \boldsymbol{\lambda},\textbf{H},\textbf{z}_1,\textbf{z}_2\right ) \nabla_{\textbf{A}_{1}} [\Psi_{\hat{\textbf{N}}}(\textbf{z}_1;\textbf{A}_1|\textbf{H},\boldsymbol{\lambda})
    ] ~  \right\},\! 
\end{align}
which is the same as \eqref{eq:gradientEstimation1}.

Similar to \eqref{eq:gradientEstimation1}, we can derive \eqref{eq:gradientEstimation2} by following steps,
\begin{align} 
    &\nabla_{\textbf{A}_{2}} \mathbb{E}_{\boldsymbol{\lambda},\textbf{H},\textbf{N},\textbf{M}} f_{\rm loss} \left ( \boldsymbol{\lambda},\textbf{H},\textbf{N}, \textbf{M} \right ) \nonumber\\
     =& \mathbb{E}_{\boldsymbol{\lambda},\textbf{H}} \int_{\textbf{z}_1,\textbf{z}_2} f_{\rm loss} \left ( \boldsymbol{\lambda},\textbf{H},\textbf{z}_1,\textbf{z}_2\right ) \Psi_{\hat{\textbf{N}}}(\textbf{z}_1;\textbf{A}_1|\textbf{H},\boldsymbol{\lambda})
    \nabla_{\textbf{A}_{2}} [\Psi_{\hat{\textbf{M}}|\hat{\textbf{N}}}(\textbf{z}_2;\textbf{A}_2|\textbf{H},\textbf{z}_1)] ~ d{\textbf{z}_1}  d{\textbf{z}_2} \nonumber\\
    =&  \mathbb{E}_{\boldsymbol{\lambda},\textbf{H},\textbf{z}_1,\textbf{z}_2} \left\{f_{\rm loss} \left ( \boldsymbol{\lambda},\textbf{H},\textbf{z}_1,\textbf{z}_2 \right )  \nabla_{\textbf{A}_{2}}[ \log \Psi_{\hat{\textbf{M}}|\hat{\textbf{N}}}(\textbf{z}_2;\textbf{A}_2|\textbf{H},\textbf{z}_1) ]\right\}. \! 
\end{align} 

\bibliographystyle{IEEEtran}
\bibliography{Reference}
\end{document}